\documentclass[aps,prb,superscriptaddress,reprint,amsmath,amssymb]{revtex4-2}
\usepackage{graphicx,color}
\usepackage{epstopdf}
\usepackage{dcolumn}
\usepackage{bm}
\usepackage{tabularx}
\usepackage{multirow}

\begin{document}
\title{Freezing of molecular rotation in a paramagnetic crystal studied by $^{31}$P NMR}

\author{D.~Opherden}
\email[Corresponding author. E-mail: ]{d.opherden@hzdr.de}
\author{F.~B\"{a}rtl}
\affiliation{Hochfeld-Magnetlabor Dresden (HLD-EMFL) and W\"{u}rzburg-Dresden Cluster of Excellence ct.qmat, Helmholtz-Zentrum Dresden-Rossendorf, 01328 Dresden, Germany}
\affiliation{Institut f\"{u}r Festk\"{o}rper- und Materialphysik, Technische Universit\"{a}t Dresden, 01062 Dresden, Germany}
\author{Sh.~Yamamoto}
\affiliation{Hochfeld-Magnetlabor Dresden (HLD-EMFL) and W\"{u}rzburg-Dresden Cluster of Excellence ct.qmat, Helmholtz-Zentrum Dresden-Rossendorf, 01328 Dresden, Germany}
\author{Z.~T.~Zhang}
\affiliation{Hochfeld-Magnetlabor Dresden (HLD-EMFL) and W\"{u}rzburg-Dresden Cluster of Excellence ct.qmat, Helmholtz-Zentrum Dresden-Rossendorf, 01328 Dresden, Germany}
\affiliation{Anhui Province Key Laboratory of Condensed Matter Physics at Extreme Conditions, High Magnetic Field Laboratory, Chinese Academy of Sciences, Hefei 230031, China}
\author{S.~Luther}
\author{S.~Molatta}
\author{J.~Wosnitza}
\affiliation{Hochfeld-Magnetlabor Dresden (HLD-EMFL) and W\"{u}rzburg-Dresden Cluster of Excellence ct.qmat, Helmholtz-Zentrum Dresden-Rossendorf, 01328 Dresden, Germany}
\affiliation{Institut f\"{u}r Festk\"{o}rper- und Materialphysik, Technische Universit\"{a}t Dresden, 01062 Dresden, Germany}

\author{M.~Baenitz}
\affiliation{Max Planck Institute for Chemical Physics of Solids, 01187 Dresden, Germany}

\author{I.~Heinmaa}
\author{R.~Stern}
\affiliation{National Institute of Chemical Physics and Biophysics, Akadeemia tee 23, 12618 Tallinn, Estonia} 

\author{C.~P.~Landee}
\affiliation{Department of Physics, Clark University, Worcester, Massachusetts 01610, USA}

\author{H.~K\"{u}hne}
\email[Corresponding author. E-mail: ]{h.kuehne@hzdr.de}
\affiliation{Hochfeld-Magnetlabor Dresden (HLD-EMFL) and W\"{u}rzburg-Dresden Cluster of Excellence ct.qmat, Helmholtz-Zentrum Dresden-Rossendorf, 01328 Dresden, Germany}

\date{\today}

\begin{abstract}

We present a detailed $^{31}$P nuclear magnetic resonance (NMR) study of the molecular rotation in the compound [Cu(pz)$_{2}$(2-HOpy)$_{2}$](PF$_{6}$)$_{2}$, where pz = C$_4$H$_4$N$_2$ and 2-HOpy = C$_5$H$_4$NHO.
Here, a freezing of the PF$_6$ rotation modes is revealed by several steplike increases of the temperature-dependent second spectral moment, with accompanying broad peaks of the longitudinal and transverse nuclear spin-relaxation rates.
An analysis based on the Bloembergen-Purcell-Pound (BPP) theory quantifies the related activation energies as $E_{a}/k_{B}$ = 250 and 1400~K.
Further, the anisotropy of the second spectral moment of the $^{31}$P absorption line was calculated for the rigid lattice, as well as in the presence of several sets of PF$_6$ reorientation modes, and is in excellent agreement with the experimental data.
Whereas the anisotropy of the frequency shift and enhancement of nuclear spin-relaxation rates is driven by the molecular rotation with respect to the dipole fields stemming from the Cu ions, the second spectral moment is determined by the intramolecular interaction of nuclear $^{19}$F and $^{31}$P moments in the presence of the distinct rotation modes.

\end{abstract}
\pacs{---}
\maketitle

\section{Introduction}
The mechanism of enhanced nuclear spin relaxation, caused by thermally activated local field fluctuations at the nuclear Larmor frequency, was first introduced in 1947 \cite{bloembergen_nuclear_1947} and elucidated in more details \cite{bloembergen_relaxation_1948} by Bloembergen, Purcell, and Pound (BPP).
Since then, the BPP mechanism of enhanced nuclear spin relaxation was observed in numerous material classes, such as cuprates~\cite{suh_spin_2000, curro_inhomogeneous_2000, mitrovic_similar_2008, baek_magnetic_2015}, iron-based superconductors~\cite{hammerath_progressive_2013, moroni_competing_2016}, spin glasses~\cite{zong_structure_2007}, low-dimensional quantum magnets~\cite{wolter_observation_2005, raffa_low-energy_1998, imai_cu_2008}, molecular magnets~\cite{borsa_nmr_2007}, organic conductors~\cite{creuzet_1h-nmr_1982, takigawa_evidence_1986}, fullerene-based superconductors~\cite{yoshinari_molecular_1993, yoshinari_molecular_1996}, materials for lithium-ion batteries~\cite{kuhn_li_2011, wilkening_microscopic_2007, kuhn_nmr_2012}, nanostructured systems~\cite{majer_nmr_2003}, and several others~\cite{ramanuja_nmr_2006, miller_nmr_1963, albert_nmr_1972, winter_ferroelectric_1992, vogelsberg_ultrafast_2017}.
Despite the diverse physical origins of the local field fluctuations in these materials, the BPP formalism provides an adequate phenomenological description of the increased nuclear spin relaxation.

As a prominent example of the related phenomenology, the coexistence and interplay of unconventional superconductivity and low-dimensional magnetism is one of the most extensively investigated topics in the research of strongly correlated electron systems.
In the iron-based and high-$T_{c}$ cuprate superconductors, low-energy spin dynamics are proposed as a key ingredient for the manifestation of superconductivity.
Until now, it is an unresolved issue under which conditions these spin fluctuations may be constructive or rather detrimental to the formation of Cooper pairs~\cite{hammerath_progressive_2013}.
In the case of glassy spin freezing, manifested as a peak of the nuclear spin-lattice relaxation rate, the behavior of the local field fluctuations can be understood as a slowing down of the characteristic electronic spin-fluctuation frequency with a Lorentzian spectral density of the fluctuations at the Larmor frequency $\omega_{0}$ \cite{hammerath_progressive_2013, moroni_competing_2016, suh_spin_2000, curro_inhomogeneous_2000, mitrovic_similar_2008, baek_magnetic_2015}.
The fluctuation rate $\tau_{c}^{-1}$, describing a thermally activated process with a distribution of activation energies $E_{a}$, indicates, e.g., glassy spin dynamics of unresolved nature in the cuprates~\cite{suh_spin_2000, curro_inhomogeneous_2000, mitrovic_similar_2008, baek_magnetic_2015}. 

Low-dimensional quantum spin systems are model materials for the study of magnetic correlations that are also present in unconventional superconductors.
In these materials, the spin-spin exchange coupling is often given by superexchange interactions.
In molecular-based materials, there is a manifold of possibilities for the occurrence of thermally activated molecular motions~\cite{wolter_observation_2005}.
These structural fluctuations may also affect the absolute values of the superexchange energies, as they relate to structural parameters of the exchange pathways.
Therefore, a detailed knowledge of the molecular fluctuation parameters is of great interest and may represent an important input for density functional theory (DFT) calculations of the exchange constants and structural parameters.

A recent nuclear magnetic resonance (NMR) study of ultrafast molecular rotation in metal-organic frameworks proposes molecular-based compounds as promising candidates for the realization of smart materials and artificial molecular machines.
Here, the utilization of molecular dynamics may lead to a tuning of the thermal, dielectric, or optical properties~\cite{vogelsberg_ultrafast_2017}.

In view of this very rich phenomenology of BPP-type local field fluctuations, and despite of several decades of research, there is a strong need for model materials that allow for a well-defined investigation of spin relaxation caused by the presence of BPP-type field fluctuations.
In particular, the study of rotation modes in single-crystalline materials allows us to probe the anisotropy of spectral properties and to perform a detailed analysis of the local field contributions at the nuclear sites. 

In the present work, we investigate molecular motions in a paramagnetic single crystal of the recently synthesized molecular-based compound CuPOF~\cite{opherden_cupof_2020} by a comprehensive NMR study of both the static and dynamic local field properties, including analysis by the BPP formalism as well as calculations of the spectral properties.

\begin{figure}[bp]
	\centering
    \includegraphics[width=.99\columnwidth]{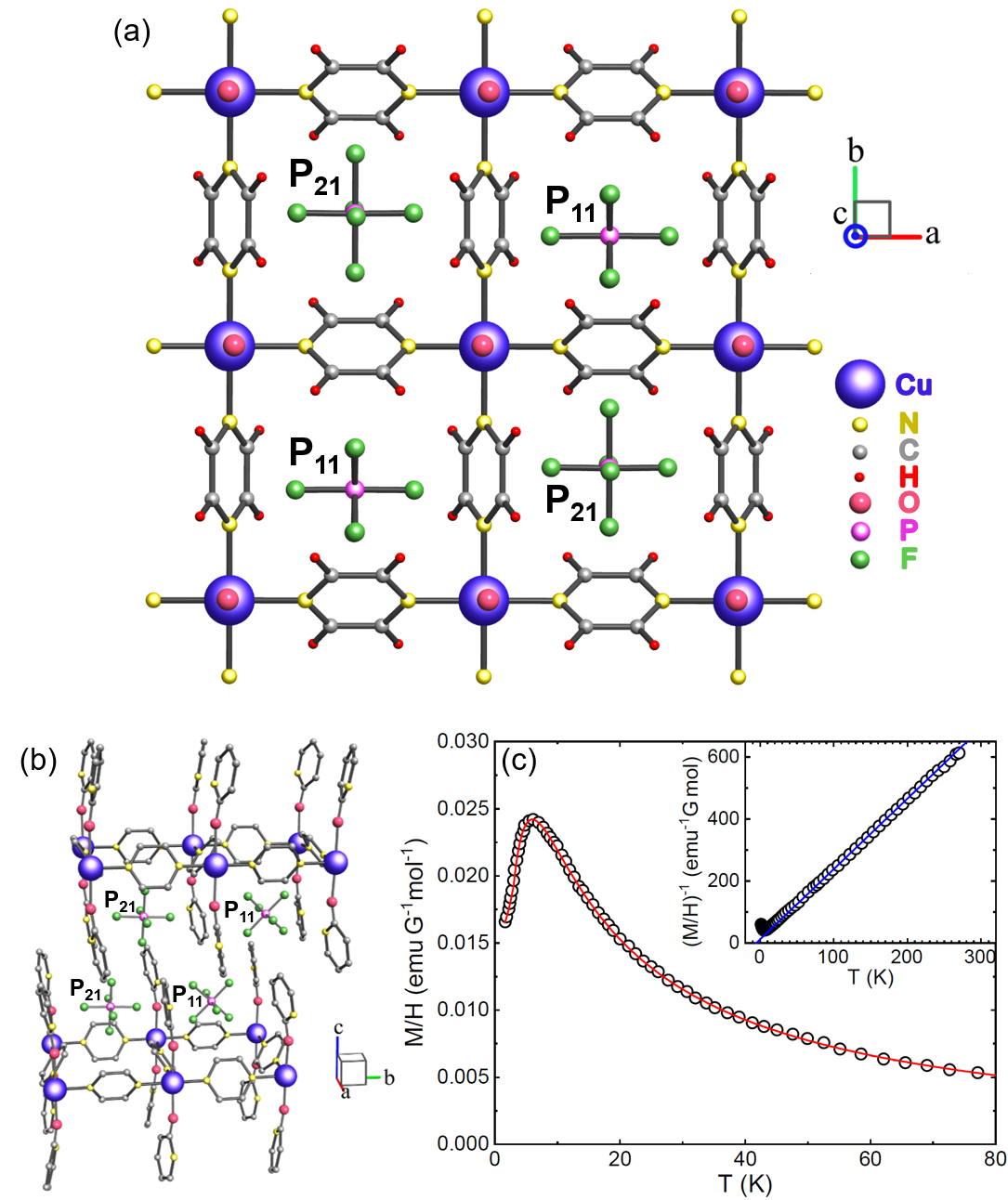}
	\caption{(a) In CuPOF, quasi-2D planes of Cu$^{2+}$ ions are linked by pyrazine molecules, forming the motif of a magnetic square lattice. Two structurally slightly inequivalent PF$_6$ molecules are located in-between the layers. (b) View on two adjacent Cu-pyrazine layers, with hydrogen atoms omitted for clarity. (c) Temperature dependence of the static susceptibility (circles). The red line shows the best fit of 2D QHAF model calculations to the magnetic susceptibility of a polycrystalline sample~\cite{opherden_cupof_2020}. The inset of (c) shows the temperature dependence of the inverse static susceptibility. The blue line denotes the best fit with a Currie-Weiss behavior.}
	\label{fig:crystal_structure}
\end{figure}

\section{Experimental}
Single-crystalline samples of the molecular-based compound [Cu(pz)$_{2}$(2-HOpy)$_{2}$](PF$_{6}$)$_{2}$ (CuPOF) were grown from solution~\cite{opherden_cupof_2020}.
Spin moments with $S=1/2$, hosted by the Cu$^{2+}$ ions, are coupled by a molecular matrix of pyrazine molecules [pz = C$_4$H$_4$N$_2$] to form quasi-two-dimensional (2D) layers with a square-lattice motif along the $ab$ plane [Fig.~\ref{fig:crystal_structure}(a)], with a nearest-neighbor intralayer exchange of $J/k_\mathrm{B}=6.80(5)$~K and an interlayer exchange of about 1~mK~\cite{opherden_cupof_2020}.
The Cu-pyrazine planes are separated by 2-pyridone molecules [2-HOpy = C$_5$H$_4$NHO] along the crystallographic $c$ axis, bridged to the Cu$^{2+}$ ions via oxygen.
These molecules not only repel the molecular planes from each other, but also cause a shift of adjacent layers by about half of an in-plane lattice period along the crystallographic $b$ axis, see Fig.~\ref{fig:crystal_structure}(b). 
The PF$_6$ anions are located in-between the layers, thus contributing to the interlayer repulsion, as well as to the effective distribution of charge density in the molecular structure.

The $^{31}$P NMR investigations of a plate-like single-crystalline sample of CuPOF, with dimensions of $4.0 \times 3.7\times 0.7$~mm$^{3}$ and a weight of about 12~mg, were performed at temperatures between 6 and 260~K, with applied magnetic fields between 2 and 7~T .
In the investigated temperature range, the correlations of the electronic moments in CuPOF can be treated as paramagnetic in good approximation.
Figure~\ref{fig:crystal_structure}(c) shows the temperature dependence of the static susceptibility of a powder sample~\cite{opherden_cupof_2020}.
Above around 40~K, the data are well described by a Curie-Weiss law with a Curie constant of 0.440(5)~emuG$^{-1}$mol$^{-1}$K and a Curie-Weiss temperature $\Theta_\mathrm{CW}= -5.2(6)$~K, indicating a small antiferromagnetic interaction, as revealed by the temperature dependence of the inverse static susceptibility in the inset of Fig.~\ref{fig:crystal_structure}(c).
Towards low temperatures, the development of the short-range correlations of the electronic moments yields a broad maximum at around 6.8~K, compare Fig.~\ref{fig:crystal_structure}(c).
This behavior was modeled with the susceptibility of a two-dimensional quantum Heisenberg antiferromagnet (2D QHAF) model with a nearest-neighbor intralayer interaction of $J/k_\mathrm{B}=6.80(5)$~K, see Ref.~\cite{opherden_cupof_2020}.
The best of the modeling process is denoted by the red line in Fig.~\ref{fig:crystal_structure}(c).

The spectra and the nuclear spin-spin relaxation time $T_2$ were recorded with a Hahn spin-echo pulse sequence with a typical $\pi/2$ pulse duration of 2.5~$\mu$s, an output power of 30~W, and waiting time  of $\tau_2 = 55$~$\mu$s (for the spectra) between the NMR radio-frequency pulses.
The experimental relaxation data of the nuclear magnetization component $M_\mathrm{xy}$ at selected temperatures are presented in the Supplemental Material (SM)~\cite{supplemental_material}, see Fig.~S3(a).
The nuclear spin-lattice relaxation time $T_1$ was recorded by using an inversion-recovery method.
The experimental relaxation data of the nuclear magnetization component $M_\mathrm{z}$ were modeled as $M_\mathrm{z} \left( \tau_1 \right)=M_0 \left[ 1-2 \exp\left( -\left( \tau_1 /T_1 \right)^{\beta} \right) \right]$, see Fig.~S1 in the SM~\cite{supplemental_material}, where $\beta$ is a stretching exponent.
We find that the stretching exponent is essentially temperature independent in the investigated temperature range, with experimental values close to unity, see Fig.~S2(b) in the SM ~\cite{supplemental_material}, indicating a uniform spin-lattice relaxation rate of the $^{31}$P nuclear ensemble.
For the temperature-dependent NMR measurements, the magnetic field was applied parallel to the crystallographic $c$ axis, i.e., perpendicular to the molecular layers.
$^{31}$P magic-angle spinning (MAS) NMR spectroscopy was performed on a polycrystalline CuPOF sample at an external field of 4.7~T with a $^{31}$P resonance frequency of
80.985~MHz and a sample spinning speed of 30~kHz.

\section{Results and discussion}

\begin{figure}[tbp]
	\centering
	\includegraphics[width=.99\columnwidth]{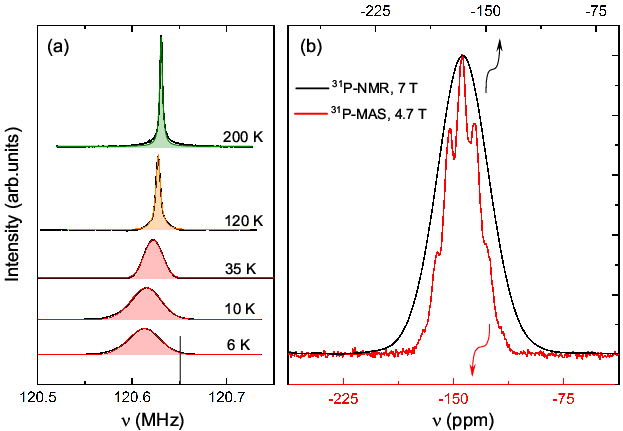}
	\caption{(a) $^{31}$P-NMR spectra of CuPOF at different temperatures and an external field of 7~T, applied parallel to the $c$ axis. The red, orange, and green shaded areas denote the modeling of the spectral lineshape by a Gaussian, Pseudo-Voigt, and Lorentzian function, respectively. The vertical line indicates the Larmor frequency for zero shift. (b) Comparison between conventional spin-echo $^{31}$P-NMR and MAS spectra at about 200~K.}
	\label{fig:spectra_NMR_MAS}
\end{figure}

Figure \ref{fig:spectra_NMR_MAS}(a) shows representative $^{31}$P-NMR spectra at selected temperatures between 6 and 200~K. Since $^{31}$P has a nuclear angular momentum of $I=1/2$, only one transition ($m_{z}=-1/2$ to $+1/2$) is observed.
Figure \ref{fig:spectra_NMR_MAS}(b) shows a comparison of the conventional spin-echo $^{31}$P-NMR spectrum and that recorded by the MAS technique at about 200~K.
Due to the cancellation of local dipole fields, by spinning the sample with a frequency of 30~kHz at the magic-angle orientation, the MAS spectrum is revealed as $J$ resolved with 7 lines, separated by a scalar spin-spin coupling of $J = 712$~Hz between the $^{19}$F and $^{31}$P nuclei~\cite{opherden_cupof_2020}.
The chemical shift of -143.2~ppm at 4.7~T and $J$ are in a good agreement with previously reported values for compounds containing PF$_{6}$ molecules~\cite{andrew_19f_1970, muetterties_structure_1959, alyea_identification_1987}.
The relative intensities of the septet are given by the binomial coefficients \cite{gutowsky_nuclear_1951}.
On the other hand, due to nuclear dipole-dipole broadening, the standard NMR spectrum is not $J$ resolved.
The small asymmetry of the NMR spectrum is attributed to the existence of two structurally slightly nonequivalent $^{31}$P sites, see Figs.~\ref{fig:crystal_structure}(a) and \ref{fig:crystal_structure}(b).
The absolute values of the MAS and NMR frequency shift differ by 22~ppm.
This is attributed to a residual dipolar contribution to the NMR shift at 200~K, which is absent in the MAS spectrum.

The full width at half maximum (FWHM) of the NMR spectrum at 200~K is less than 5~kHz (41~ppm at 7~T), indicating a high homogeneity of the single-crystalline CuPOF sample.
As shown in Fig. \ref{fig:spectra_NMR_MAS}(a), the linewidth increases significantly upon cooling.
Since the $^{31}$P-NMR spectrum is slightly asymmetric, the second spectral moment $M_2$, the square root of which is proportional to the FWHM, is used to characterize the spectral width.
The $n$th spectral moment is defined as $M_n= \int \left( \omega - \omega_{0} \right)^{n}f(\omega) d \omega$, where the resonance curve is described by a normalized function $f (\omega)$ with a maximum at the frequency $\omega_{0}$~\cite{abragam_principles_1961}.
The temperature dependence of the $^{31}$P second spectral moment is shown in Fig. \ref{fig:BPP}(a).
Instead of a simple Curie-type behavior, i.e., a monotonic increase of $\sqrt{M_2}$ toward low temperatures~\cite{zhang_defect_2017}, which would denote a continuously growing width of the magnetic dipole-field distribution stemming from the Cu$^{2+}$ moments, the temperature dependence of the linewidth shows several steplike increases.
At temperatures below about 60~K, the $^{31}$P-NMR spectrum resembles a Gaussian lineshape, as exemplified by the red shaded areas in Fig.~\ref{fig:spectra_NMR_MAS}(a).
With increasing temperatures, the spectral lineshape gradually changes, and a purely Lorentzian form is observed above around 145~K, compare the green shaded area in Fig.~\ref{fig:spectra_NMR_MAS}(a).
At intermediate temperatures, between about 60 and 145~K, the $^{31}$P-NMR spectrum can be described by a superposition of Lorentzian and Gaussian functions, displayed as the orange shaded area in Fig.~\ref{fig:spectra_NMR_MAS}(a).
Furthermore, the temperature-dependent spin-spin and spin-lattice relaxation rates, $1/T_2$ and $1/T_1$, display several broad maxima, compare Figs.~\ref{fig:BPP}(b) and \ref{fig:BPP}(c).
The increases of $1/T_2$ coincide with temperature regimes for which the steplike changes of the linewidth are found, whereas the related maxima of $1/T_1$ are observed at higher temperatures.
The broad maxima of the temperature-dependent $1/T_1$ and $1/T_2$ rates, as well as the steplike behavior of the linewidth, are ascribed to a motional narrowing process, as described further below.

\subsection{Nuclear relaxation and BPP phenomenology}

According to the BPP model~\cite{bloembergen_nuclear_1947,bloembergen_relaxation_1948}, a time-dependent local magnetic field $\vec{h}(t)$, stemming from either nuclear or electronic moments, with a characteristic fluctuation frequency close to that of the nuclear Larmor frequency $\omega_{0}$, represents a mechanism of nuclear spin-lattice relaxation:
\begin{equation}
\label{form:BPP}
\frac{1}{T_1}=\gamma^{2}\langle{h_{\bot}^{2}}\rangle\frac{\tau_{c}}{1+\left(\omega_{0}\tau_{c}\right)^{2}}.
\end{equation}
Here, ${\gamma}$ is the nuclear gyromagnetic ratio, and ${h_{\bot}}$ is the perpendicular component of $\vec{h}(t)$ with a mean-square amplitude of  $\langle{h_{\bot}^{2}}\rangle$.
In the case of liquids, where the effect was first observed, $\tau_{c}$ is a correlation time associated with local Brownian motion, whereas in gases, $\tau_{c}$ is the average time between molecular collisions. Gutowsky and Pake showed that the same approach can be applied to the study of atomic motion in solids, by treating $\tau_{c}$ as an average time between jumps from one atomic site to another \cite{gutowsky_structural_1950}. Typically, the temperature dependence of the correlation time can be described as a thermally activated process:
\begin{equation}
\label{form:tau_c}
\tau_{c}=\tau_{0} \exp\left(\frac{E_{a}}{k_{B}T}\right),
\end{equation}
with the activation energy $E_{a}$ and the infinite-temperature correlation time $\tau_{0}$.

\begin{figure}[tbp]
	\centering
	\includegraphics[width=0.99\columnwidth]{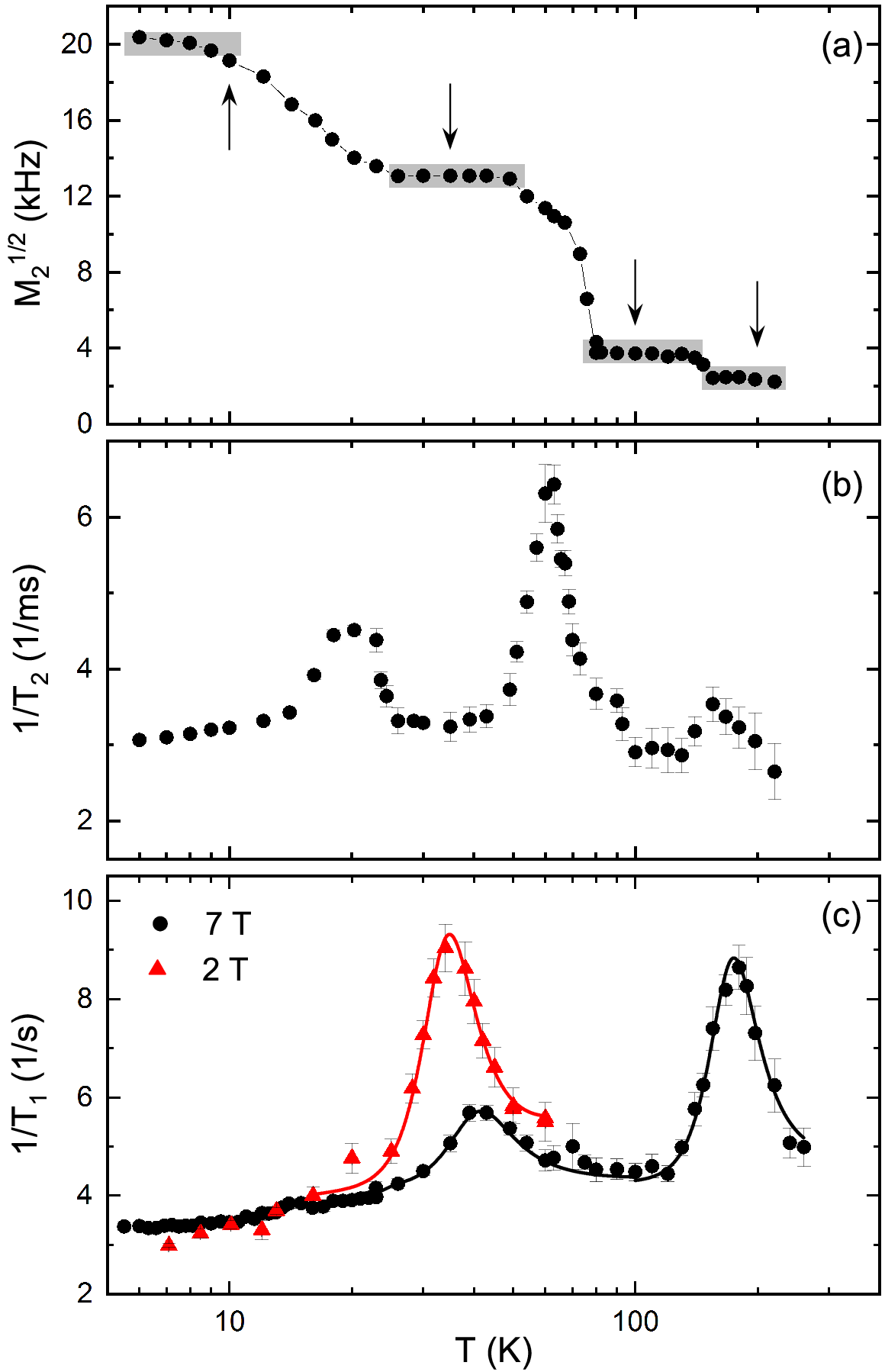}
	\caption{(a) Temperature dependence of the square root of the $^{31}$P second spectral moment at 7~T. The vertical arrows indicate the temperatures at which the angular-dependent $^{31}$P NMR spectra were recorded, see Figs.~\ref{fig:ang_dep-1st-moment} and \ref{fig:ang_dep}. (b) Temperature dependence of the $^{31}$P nuclear spin-spin relaxation rate at 7~T. (c) Temperature dependence of the $^{31}$P nuclear spin-lattice relaxation rate at 7 and 2~T. The black (red) lines represent fits with the BPP model to the 7~T (2~T) data.}
	\label{fig:BPP}
\end{figure}

In most experiments, with only a few exceptions \cite{gutowsky_structural_1950, miller_nmr_1963, albert_nmr_1972, winter_ferroelectric_1992}, only one broad maximum of the spin-lattice relaxation rate is observed, eventually accompanied by a steplike change of the temperature-dependent second moment.
However, in the present case of CuPOF, each $1/T_1$ peak and related constant regime of the temperature-dependent second moment can be attributed to a different set of characteristic rotational modes of the PF$_6$ molecules.
The temperature-dependent $1/T_1$ data at 7~T was modeled using Eqs. (\ref{form:BPP}) and (\ref{form:tau_c}) as well as a weak linear contribution.
For the two distinguishable maxima of $1/T_1$, two separate sets of BPP parameters were used, compare Fig.~\ref{fig:BPP}(c).
The perpendicular component of the local fluctuating field is determined as $(\langle{h_{\bot,l}^{2}}\rangle)^{1/2}=1.1$~mT for the low-temperature peak, while for the high-temperature peak, we find $(\langle{h_{\bot,h}^{2}}\rangle)^{1/2}=1.8$~mT.
Since these values are very similar, we conclude that both processes are related to motional modes of the same physical object.
The activation energies and related correlation times are $E_{a,l}/k_{B}=250$ K and $\tau_{0,l}=20$~ps for the low-temperature process, and $E_{a,h}/k_{B}=1400$~K and $\tau_{0,h}=3$~ps for the high-temperature one, respectively.
Both the activation energies and the correlation times are in the parameter range observed for other compounds with a BPP-type enhancement of $1/T_1$ by thermally activated reorientation modes \cite{albert_nmr_1972, gutowsky_pulsed_1973, ripmeester_molecular_1979}.

In order to further test the validity of the BBP model in the present case, additional $1/T_1$ measurements were performed at 2~T.
According to the BBP theory, a decrease of the magnetic field and the corresponding NMR frequency leads to an increase of the peak amplitude and a temperature downshift of the peak position.
The parameters $\langle{h_{\bot}^{2}}\rangle$, $\tau_{0}$, and $E_{a}$ of the molecular rotation remain unchanged by the variation of the field amplitude.
These predictions are fully compatible with our experimental results.
The BPP modeling of the $1/T_1$ data at 2~T gives $(\langle{h_{\bot,l}^{2}}\rangle)^{1/2}=1.0$~mT, $\tau_{0,l}=20$~ps, and $E_{a,l}/k_{B}=255$~K.

The change of molecular reorientation modes also yields three broad maxima of $1/T_2$ at around 20, 60, and 150~K, coinciding with the temperature regimes of the steplike growth of the second moment, compare Figs.~\ref{fig:BPP}(a) and \ref{fig:BPP}(b).
In general, the $T_2$ relaxation can be caused by several mechanisms, such as nuclear spin-spin coupling (direct or indirect), or by slow longitudinal local field fluctuations.
Whereas the first mechanism may yield complex transverse relaxation depending on the details of the internuclear interactions~\cite{pennington_nmr_1989}, the latter mechanism, commonly referred to as the Redfield contribution, usually leads to a solely exponential decay with a time constant coupled to the spin-lattice relaxation.
For the whole investigated temperature range, we find the spin-echo decay to be purely exponential, modulated by a weak oscillatory component, stemming from the spin-spin coupling between the $^{31}$P and $^{19}$F nuclei.
This oscillatory component yields a frequency of about 790(70)~Hz, which is in good agreement with $J = 712$~Hz, determined by the MAS-NMR spectroscopy, see Fig.~\ref{fig:spectra_NMR_MAS}(b).
The observed exponential decay is in contrast to, e.g., a more complex temperature evolution of the $^{13}$C spin-echo decay, reported for molecular reorientations in the fullerene-based superconductor K$_3$C$_{60}$~\cite{yoshinari_molecular_1993}.

Considering the purely exponential transverse relaxation in CuPOF, but also that $1/T_2$ quantitatively clearly exceeds $\sqrt{M_2}$ at most temperatures in the parameter regime of our study, the main mechanism of the $T_2$ relaxation is, most likely, not given by the internuclear interaction within the PF$_{6}$ molecules, but is rather due to longitudinal local field fluctuations, originating from the dipolar fields of the Cu$^{2+}$ electronic moments.
A comprehensive quantitative description of the $T_2$ relaxation process in CuPOF is, however, beyond the scope of the present work. 
	
In the rather complex molecular-based structure of CuPOF, there are several candidates for yielding molecular reorientation modes, namely the pyrazine molecules, the 2-pyridone molecules, and the PF$_{6}$ anions.
The pyrazine molecules are bridging the Cu$^{2+}$ ions, constituting a superexchange pathway for the electronic moments, so that a rotational motion about the axis that links nearest copper neighbors might influence the related Heisenberg exchange coupling $J$~\cite{opherden_cupof_2020}.
However, both the pyrazine as well as the 2-pyridon molecules may be expected to yield only one characteristic set of rotational modes, as they are correspondingly chemically bound to the layered structure.

Since the PF$_{6}$ molecules are not bound to any other part of the molecular structure, they have the highest degree of freedom, which makes them the most likely candidates to yield complex motional modes.
There are a few NMR studies of the rotational motions of the PF$_{6}$ anions in alkali hexafluorophosphates to compare with~\cite{miller_nmr_1963,albert_nmr_1972,gutowsky_pulsed_1973}.

\subsection{Anisotropy of the spectral moments}

\begin{figure}[tbp]
	\centering
	\includegraphics[width=0.99\columnwidth]{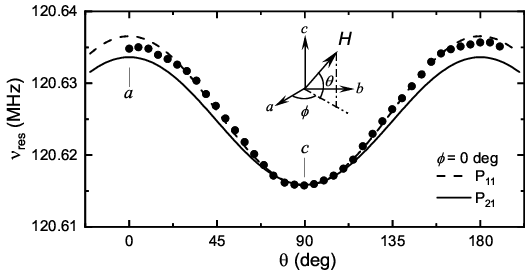}
	\caption{Angular dependence of the $^{31}$P-NMR frequency shift (full circles), recorded for a field rotation in the crystallographic $ac$ plane at 7~T and 10~K. For both the P$_{11}$ and P$_{21}$ sites, the experimental data are compared to the calculated anisotropy of the dipolar-field contribution, stemming from the Cu$^{2+}$ ions. The sketch in the inset defines the polar and azimuthal angles $\theta$ and $\phi$, respectively. The case $\phi =0$ denotes the rotation in the $ac$ plane.}
	\label{fig:ang_dep-1st-moment}
\end{figure}

In order to obtain further quantitative understanding of the rotation of the PF$_{6}$ molecules and the related motional narrowing of the NMR spectra, measurements of the angular-dependent $^{31}$P resonance shift were performed at 10, 35, 100, and 200~K, selecting characteristic regions of the stepwise temperature dependence of $\sqrt{M_2}$, see Fig.~\ref{fig:BPP}(a).
Numerical estimates of the anisotropic NMR shift for the two structurally slightly non-equivalent $^{31}$P sites were performed by a summation of the dipole fields from the localized Cu$^{2+}$ electronic moments in a volume of (100~\AA)$^3$.
The results of these calculations are in very good agreement with the experimental data, see Fig.~\ref{fig:ang_dep-1st-moment}, and reveal that the full magnetic moment is, in fact, localized at the Cu$^{2+}$ sites.
The isotropic chemical-shift contribution is determined as -168~ppm.
  
\begin{figure}[bp]
	\centering
	\includegraphics[width=0.99\columnwidth]{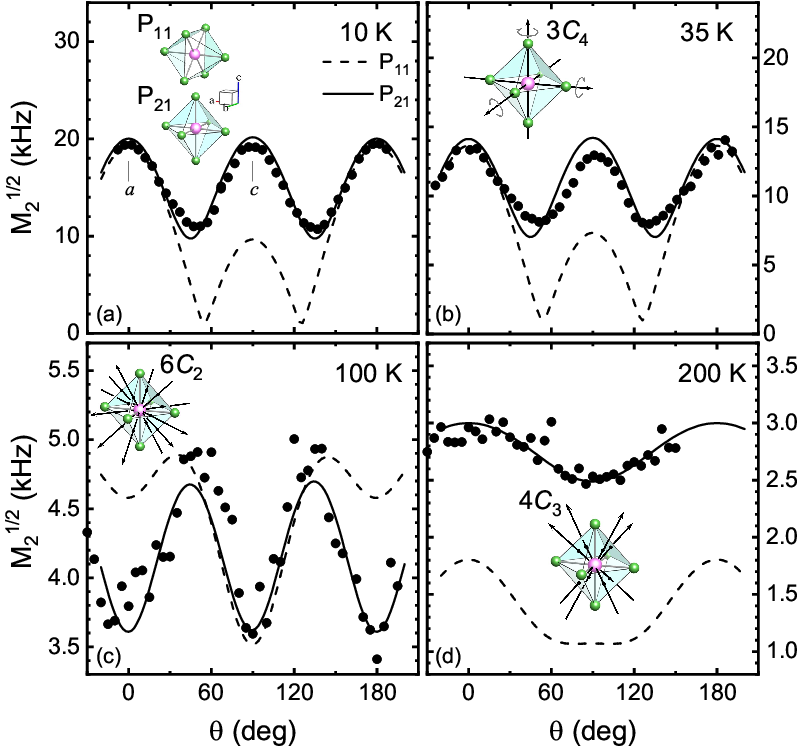}
	\caption{Angular dependence (with $H \, \| \,a$ or $ \| \, c$ corresponding to $\theta = 0$ or 90 degrees, respectively) of square root of the $^{31}$P second spectral moment at (a) 10, (b) 35, (c) 100, and (d) 200~K, compared to calculations for the nuclear P$_{11}$ and P$_{21}$ sites for (a) a rigid lattice and in the presence of PF$_6$ molecular reorientations around the symmetry axes (b) 3$C_4$, (c) 6$C_2$, and (d) 4$C_3$. The inset in (a) shows a sketch of the local P$_{11}$ and P$_{21}$ symmetry with respect to the in-plane ($H \, \| \,a$, horizontal) and out-of-plane ($H \, \| \, c$, vertical) field orientation. The insets in (b), (c), and (d) show sketches of the molecular symmetry axes, around which the PF$_6$ molecular reorientations occur.}
	\label{fig:ang_dep}
\end{figure}

At all temperatures, the anisotropy of the frequency shift is much smaller than the spectroscopic linewidth.
Furthermore, by lowering the temperature from one step of $\sqrt{M_2}$ to another, the periodicity of the $\sqrt{M_2}$ anisotropy changes, see Fig.~\ref{fig:ang_dep}.
In contrast, a $\pi$-periodicity of the frequency shift is observed at all temperatures.
The anisotropy amplitude scales with the magnetic dipole-field distribution stemming from the Cu$^{2+}$ moments.
These findings strongly suggest that the formation of the first ($\nu_{res}$) and the second ($M_2$) spectral moments are determined by different physical mechanisms. 

For a further quantitative analysis of the experimental data, we calculated the second spectral moment.
According to the Van Vleck theory of the nuclear-resonance linewidth in a rigid lattice~\cite{van_vleck_dipolar_1948}, the second moment $M_{2}^{I}$ of the resonance absorption of the nuclear species $j$ with angular momentum $I$ and gyromagnetic ratio $\gamma_{I}$ can be written as
\begin{equation}
\label{eq:M2}
\begin{aligned}
	M_{2}^{I}=\left(\Delta\omega\right)^{2}=\frac{3}{4}I\left(I+1\right)\sum_{k}\left[\gamma_{I}^{2}\hbar\frac{\left(1-3\cos^{2}\theta_{jk}\right)}{r_{jk}^{3}}\right]^{2}\\
  +\frac{1}{3}S\left(S+1\right)\sum_{k'}\left[J_{IS}+\gamma_{I}\gamma_{S}\hbar\frac{\left(1-3\cos^{2}\theta_{jk'}\right)}{r_{jk'}^{3}}\right]^{2},
\end{aligned}
\end{equation}
where $\theta_{jk}$ is the angle between the position vector $\vec{r}_{jk}$ of the magnetic moments $j$ and $k$ and the magnetic field, and $J_{IS}$ is the scalar coupling between unlike spins. The first term of Eq.~(\ref{eq:M2}) represents the broadening by like nuclear moments, whereas the second term of Eq.~(\ref{eq:M2}) accounts for all other magnetic moments $S^{k'}\!.$ The line broadening caused by different types of spin systems, $I^{k}$ and $S^{k'}\!,$ is additive, and the broadening by like spins is $9/4$ times more efficient than by unlike spins.

Since electronic magnetic moments are about 1000 times larger than nuclear magnetic moments (depending on the isotope), we first consider the line broadening due to dipole fields stemming from the Cu$^{2+}$ ions.
Our calculations of the anisotropic dipole broadening using the second term of Eq.(\ref{eq:M2}) gives magnitudes of 46-156~kHz, where 46~kHz results for an external magnetic field perpendicular to the $ab$ planes.
The calculated values significantly exceed the experimentally determined second moment, e.g., $\sqrt{M_2} = 19$~kHz at 10~K for the out-of-plane orientation.
The reason for this discrepancy is found by considering the exchange interactions between the Cu$^{2+}$ moments, contributing strongly to the fourth spectral moment, which leads to a narrowing of the absorption line \cite{van_vleck_dipolar_1948}.
The effective linewidth can be estimated as $\left(\Delta\omega\right)^{2}_{ \mathrm{Cu},eff}=\left(\Delta\omega\right)^{2}_{\mathrm{Cu}}h \left(\Delta\omega\right)_{obs} /A$, where $\left(\Delta\omega\right)^{2}_{\mathrm{Cu}}$ is the calculated second moment of the electronic copper moments, $\left(\Delta\omega\right)_{obs}$ is the experimentally observed width and $A/h$ is the exchange frequency \cite{bloembergen_fine_1950}.
In the case of CuPOF, $A/h \approx 4\cdot10^{11}$~Hz is estimated from the known Heisenberg exchange value of $J/k_{B} = 6.80(5)$~K~\cite{opherden_cupof_2020}.
Thus, the effective second moment contribution of the copper ions for an out-of-plane orientation of the external magnetic field is only about $\sqrt{M_2} = 25$~Hz.
In consequence, whereas the Cu$^{2+}$ moments mainly determine the first spectral moment of the $^{31}$P absorption line, they yield no significant contribution to the second moment due to the exchange narrowing.

Since the second moment is inversely proportional to the sixth order of the distance between the interacting magnetic moments, the nuclear intramolecular broadening of PF$_{6}$ is examined as the next possible mechanism of the $^{31}$P line broadening.
As mentioned, there are two structurally non-equivalent PF$_{6}$ molecules in the structure, see Figs.~\ref{fig:crystal_structure}(a) and \ref{fig:crystal_structure}(b), giving rise to different second-moment anisotropies for the corresponding sites P$_{11}$ and P$_{21}$.
Using the second term of Eq.~(\ref{eq:M2}), where $J_{IS} = J_{PF} = 712$~Hz, as determined by the MAS NMR experiments [see Fig.~\ref{fig:spectra_NMR_MAS}(b)], the calculated anisotropy for the P$_{21}$ site agrees very well with the experimental data, compare Fig.~\ref{fig:ang_dep}(a).
Due to almost identical first-moment anisotropies of the sites P$_{11}$ and P$_{21}$, see Fig.~\ref{fig:ang_dep-1st-moment}, the broader P$_{21}$ spectrum determines the experimentally observed anisotropy of the second moment.

The second-moment contribution from the dipolar broadening by the $^{13}$C, $^{14}$N, $^{15}$N, $^{63}$Cu, and $^{65}$Cu nuclei, as well as the intermolecular broadening by the $^{31}$P and $^{19}$F nuclei, was evaluated by the second term of Eq.~(\ref{eq:M2}).
The total calculated contribution yields only $\sqrt{M_2} = 26$~Hz, and is neglected in the following.

\begin{table}[tbp]
	\centering
	\caption{Comparison between the calculated and experimentally recorded second moment $\sqrt{M_2}$, for the rigid lattice and rotations around twofold, threefold, and fourfold symmetry axes. The mean-square amplitude of the transverse local magnetic-field fluctuations, $\langle{h_{\bot}^{2}}\rangle$, the correlation time at infinite temperature $\tau_{0}$, and the activation energy $E_{a}$, obtained from the BPP analysis of the $^{31}P$ spin-lattice relaxation maxima, are presented for PF$_{6}$ molecular reorientations about $6C_2$ and $4C_3$. In all cases, the magnetic field is parallel to the $c$ axis.}
	\renewcommand{\arraystretch}{1.2}
	\begin{tabularx}{\columnwidth}{cc|ccccc}
		\hline \hline
		\multicolumn{2}{c|}{Calculation} & \multicolumn{5}{c}{Experimental}\\
		\hline
		\multirow{2}{*}{Rotational mode} & $\sqrt{M_{2}}$ & T & $\sqrt{M_2}$ & $(\langle{h_{\bot}^{2}}\rangle)^{1/2}$ & $E_{a}/k_{B}$ & $\tau_{0}$\\
		& (kHz) &  (K) & (kHz) & (mT) & (K) & (ps) \\
		\hline 
		None, rigid lattice & 20.16  & 10 & 19.14 &  &  &  \\
		3C$_{4}$ & 14.21  & 35 & 12.95 & 1.1 & 250 & 20 \\
		6C$_{2}$ & 3.62  & 100 & 3.59 & - & - & - \\
		4C$_{3}$ & 2.49  & 200 & 2.53 & 1.8 & 1400 & 3 \\
		\hline \hline 
	\end{tabularx}
	\label{tab: comparison}
\end{table}

All of the above calculations are performed for a rigid lattice, i.e., under the assumption that the lengths and orientations of the vectors describing the relative positions of the magnetic moments are time independent.
A different situation appears for liquids and gases, as well as for atomic diffusion in solids.
In these cases, due to the rapid relative motion of magnetic moments, the resulting local fields at the nuclear sites fluctuate in time, and only their average value is observed experimentally.
This mechanism is known as motional narrowing, and can be observed under the condition that the average is taken over a time period which is long in comparison to the characteristic time scale of the fluctuations.
The criterion for motional narrowing is $\sqrt[]{\left(\Delta\omega_{0}\right)^{2}}\cdot\tau_{c}\ll1$, where $\left(\Delta\omega_{0}\right)^{2}$ is the second moment in the rigid lattice, described by Eq.~(\ref{eq:M2}), and $\tau_{c}$ is the correlation time characterizing the rate of the local field fluctuations.
In the present case of CuPOF, this criterion is satisfied, since $\tau_{c}$ is of the order of a few ps, whereas $\sqrt[]{\left(\Delta\omega_{0}\right)^{2}}$ is of the order of a few ten kHz, see Table~\ref{tab: comparison}.

Since the second moment of the $^{31}$P resonance is mainly determined by the interaction with the six neighboring $^{19}$F nuclei of the same PF$_{6}$ molecule, it is reasonable to consider that the motional narrowing is related to reorientation modes of the PF$_{6}$ molecule itself.
These molecules form almost perfect octahedral structures, the simplest and foremost motion of which are rotations around well-defined symmetry axes.
The rotation about the twofold, threefold, and fourfold (C$_{n}$, $n = 2$, 3, 4) symmetry axes narrows the $^{31}$P line.
In order to account for the molecular rotation, the angular-dependent coefficient $(1-3\cos^{2}\theta_{jk})$ in Eq.~(\ref{eq:M2}) needs to be replaced by its average over all possible angles $\theta_{jk}$~\cite{abragam_principles_1961}.
According to the theorem for spherical harmonics,
\begin{equation}
\label{eq:M2_coeff_rotation}
	\overline{\left(3\cos^{2}\theta_{jk}-1\right)} = \frac{1}{2}\left(3\cos^{2}\gamma-1\right)\left(3\cos^{2}\theta'-1\right),
\end{equation}
where $\gamma$ and $\theta'$ are the angles between the symmetry-axis direction vector and $\vec{r}_{jk}$ or the externally applied magnetic field, respectively.
The angular dependence of the $^{31}$P-NMR second moment $M_{2}$ of the P$_{11}$ and P$_{21}$ sites was calculated in the presence of PF$_6$ molecular reorientations around the symmetry axes $6C_2$, $4C_3$, and $3C_4$, using Eqs.~(\ref{eq:M2}) and (\ref{eq:M2_coeff_rotation}).
The angular dependences for each rotational PF$_6$ motion are compared to the experimental values of $\sqrt{M_2}$ at 35~K ($3C_4$), 100~K ($6C_2$), and 200~K ($4C_3$), as shown in Figs.~\ref{fig:ang_dep}(b)-\ref{fig:ang_dep}(d).
An excellent agreement between the calculations for the phosphorous site P$_{21}$ and the experimental data is observed.
For the given rotation in the $ac$ plane, the different crystallographic orientations of the PF$_6$ molecules of the sites P$_{11}$ and P$_{21}$, as depicted in the inset of Fig.~\ref{fig:ang_dep}(a), result in quantitative differences of the calculated second moment anisotropies for the respective sites, depending on the symmetry axes of the given molecular rotation mode. Note that the calculations of $M_2$ do not include any free parameter. 

As the only small deviation, at~100~K, the experimentally determined $\sqrt{M_2}$ mostly follows the calculations for the P$_{21}$ site, although a broader line is expected for the P$_{11}$ site.
Nevertheless, the excellent overall agreement between the experimental and calculated second moment allows assignment of the PF$_6$ molecular reorientation modes to the respective temperature regimes of constant $\sqrt{M_2}$.
The motional modes in the different temperature regimes can be described as depicted by the insets in Figs.~\ref{fig:ang_dep}(b)-\ref{fig:ang_dep}(d).
At all temperatures, the same type of the PF$_6$ molecular orientations for both inequivalent PF$_6$ molecules, yielding the P$_{11}$ and P$_{21}$ sites, are expected.

\subsection{Temperature-dependent evolution of the PF$_6$ molecular reorientations}

At temperatures above 155~K, all PF$_{6}$ molecules are rotating around the four threefold symmetry axes $4C_3$, which pass through the centers of the octahedron surfaces.
These molecular reorientations yield a constant linewidth of the $^{31}$P-NMR spectrum with $\sqrt{M_2}= 2.5$~kHz for the out-of-plane field orientation.
With decreasing temperature, these molecular reorientation modes are continuously freezing out, as is manifested by the broad high-temperature maximum of the $^{31}$P spin-lattice relaxation rate, see Fig.~\ref{fig:BPP}(c).
By lowering the temperature below 155 K, a steplike increase of the linewidth occurs with a concomitant broad maximum of $1/T_2$ [Figs.~\ref{fig:BPP}(a) and \ref{fig:BPP}(b)].
At temperatures between about 130 and 80~K, the molecular reorientation takes place with respect to the six twofold symmetry axes $6C_2$, which pass through the centers of the octahedron edges.
This type of PF$_{6}$ molecular rotations manifests as a plateau of the temperature-dependent linewidth with $\sqrt{M_2}$ of about 3.6~kHz for $H \| c$.
Although these motional modes are freezing out with further decrease of temperature, no distinct maximum of the $^{31}$P spin-lattice relaxation rate, associated with a slowing of this specific molecular reorientation, is observed.
The absence of this $1/T_1$ peak could either be caused by a significant and abrupt change of the molecular reorientation frequency, or by a cancellation of the local field fluctuations due to the anisotropy of the geometrical form factors, the anisotropy and magnitude of which determine the sensitivity of the spin-lattice relaxation to the local field dynamics.
By lowering the temperature below about 80~K, a further stepwise increase of the linewidth is observed and accompanied by an enhancement of the $^{31}$P spin-spin relaxation rate.
A broad maximum of $1/T_2$ occurs at 60~K, whereas below around 50~K, another plateau of $M_2$ and $1/T_2$ is reached, which is associated with rotations of the PF$_{6}$ molecules around the three fourfold symmetry axes $3C_4$, which coincide with the space diagonals of the octahedron.
The freezing of these molecular motions yields another BPP peak in the $^{31}$P spin-lattice relaxation rate, see Fig.~\ref{fig:BPP}(c).
Below about 26~K, an increase of the linewidth is observed, accompanied by a broad maximum of the $^{31}$P spin-spin relaxation.
Finally, at around 10~K, all PF$_{6}$ reorientation modes are frozen out and the rigid-lattice condition is reached.
A molecule with several equivalent orientations with respect to the symmetry axis, separated by energy barriers, may flip between these orientations with a given frequency.
In case of more than two such equivalent orientations of the molecule, the calculation of the second moment gives the same result for an averaging over a discrete number of possible orientations or for a classical description of the rotation, i.e., averaging over a continuous set of orientations~\cite{abragam_principles_1961}.
A summary of all experimental and calculated characteristic parameters is presented in Table~\ref{tab: comparison}.

\section{Conclusions}
In summary, we used a combined approach by probing static and dynamic local-magnetic-field properties by means of NMR spectroscopy, as well as molecular-motion models, in order to investigate the freezing of molecular rotation modes in a paramagnetic crystal CuPOF.
By analyzing the temperature-dependent relaxation rates, the activation energies $E_{a,l}/k_{B}=250$~K and $E_{a,h}/k_{B}=1400$~K, with the corresponding correlation times $\tau_{0,l}=20$~ps and $\tau_{0,h}=3$~ps of the molecular reorientations, were determined.
The angular-dependent frequency shift and second moment, compared with our calculations, reveal the origin of the low-frequency local field fluctuations.
The excellent agreement between the calculations and experimental data allows for a well-defined investigation and identification of the temperature-dependent evolution of the PF$_{6}$ molecular rotation modes.
Perspectively, the knowledge of the microscopic environment of the Cu$^{2+}$ ions and its variation with temperature can be used as an important input for detailed DFT calculations of the electronic properties in CuPOF.
The presented approach can be used for a broad range of similar molecular-based compounds with localized magnetic moments, and, thus, opens new possibilities for the exploration of molecular rotational modes in paramagnetic single crystals.
\begin{acknowledgments}
We appreciate helpful discussions with Prof. Mark Turnbull.
We acknowledge support from the Deutsche Forschungsgemeinschaft (DFG) through the SFB 1143, the GRK 1621, and the W\"{u}rzburg-Dresden Cluster of Excellence on Complexity and Topology in Quantum Matter--$ct.qmat$ (EXC 2147, Project No.\ 390858490), as well as the support of the HLD at HZDR, a member of the European Magnetic Field Laboratory (EMFL).
Z.T.Z. was supported by the National Natural Science Foundation of China (Grant No. 11304321) and by the International Postdoctoral Exchange Fellowship Program 2013 (Grant No. 20130025).
I.H. and R. S. were supported by the European Regional Development Fund (Grant No. TK134), and by the Estonian Research Council (PRG4, IUT23-7).
\end{acknowledgments}

\bibliographystyle{unsrt}

\newpage
\clearpage

\setcounter{figure}{0}
\renewcommand{\thefigure}{S\arabic{figure}}%
\setcounter{equation}{0}
\renewcommand{\theequation}{S\arabic{equation}}%

	\section{Supplemental Material}

\begin{figure}[!h]
	\centering
	\includegraphics[width=0.95\linewidth]{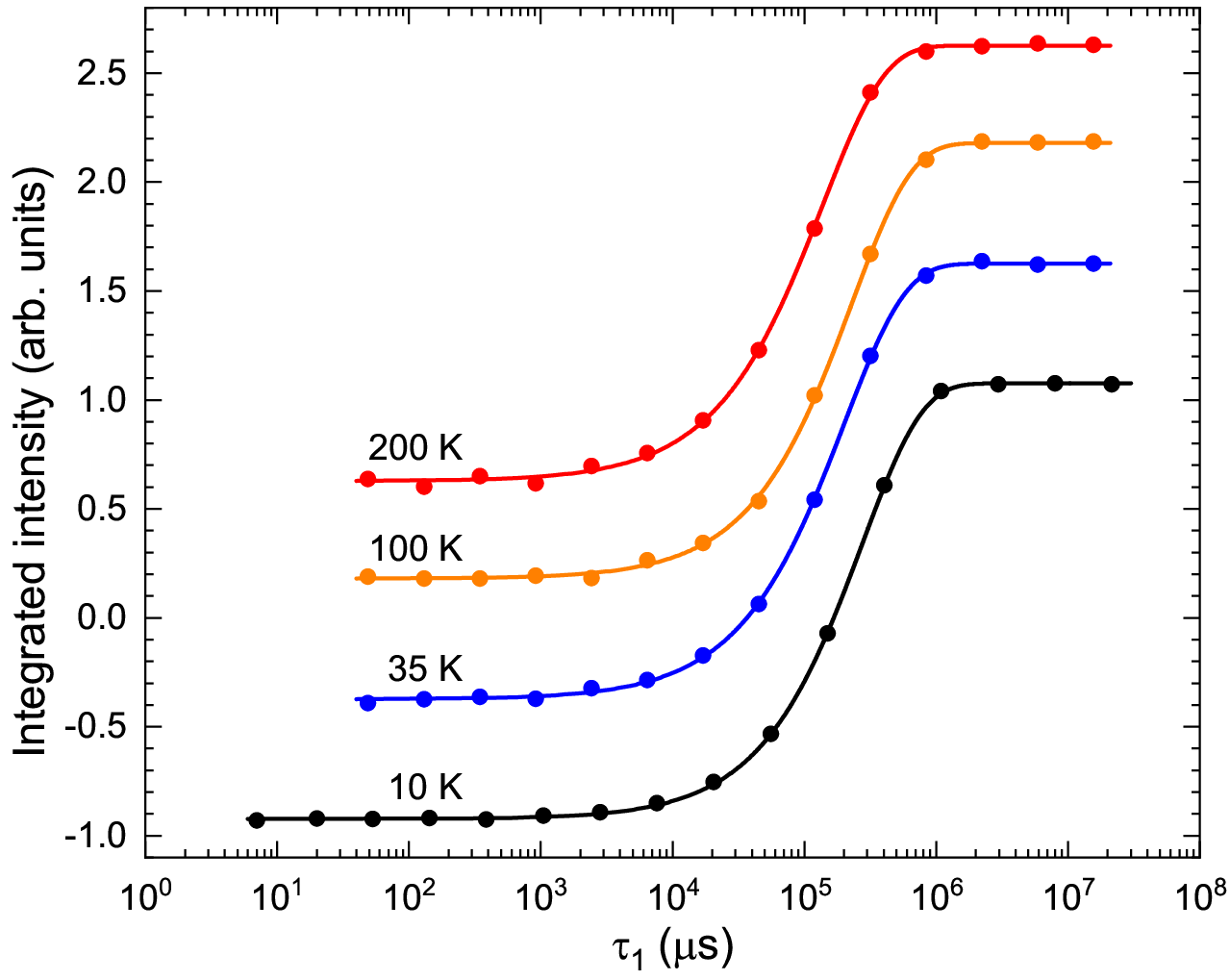}
	\caption{Experimental relaxation data of the $^{31}$P nuclear magnetization component $M_\mathrm{z}$ (circles) at 10, 35, 100, and 200 K, modeled by Eq.~(\ref{eq:T1}) (solid lines). The vertical scale corresponds to the data recorded at 10~K. The curves at higher temperatures are shifted vertically for clarity.}
	\label{fig:T1_curves}
\end{figure}

\begin{figure}[b]
	\centering
	\includegraphics[width=0.95\linewidth]{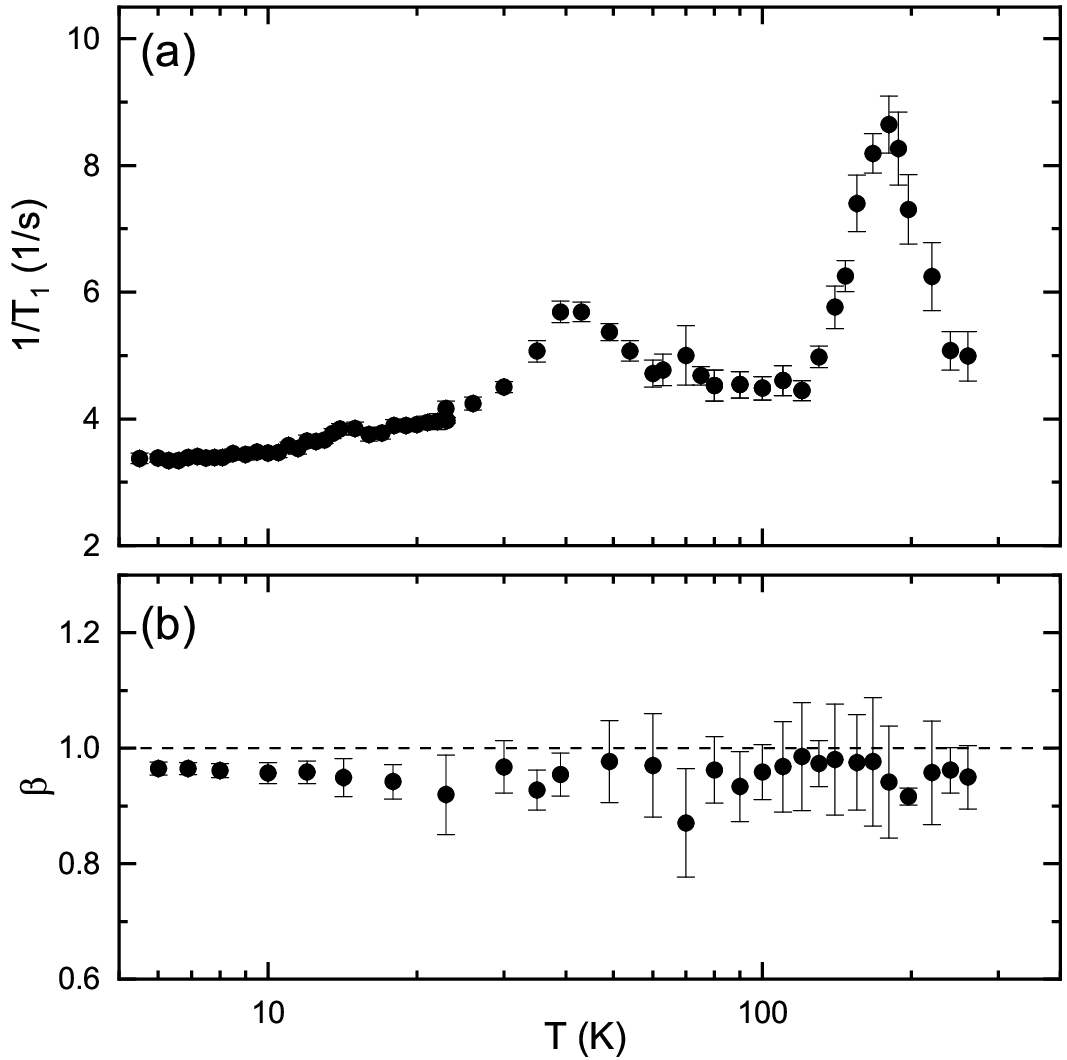}
	\caption{Temperature dependence of (a) the $^{31}$P nuclear spin-lattice relaxation rate and (b) the corresponding stretching exponent.}
	\label{fig:1overT1_and_beta}
\end{figure}

\textit{The nuclear spin-lattice relaxation time $T_1$}. The nuclear spin-lattice relaxation time $T_1$ was recorded by using an inversion-recovery method.
The experimental relaxation data of the nuclear magnetization component $M_\mathrm{z}$ recorded at 10, 35, 100, and 200~K are presented in Fig.~\ref{fig:T1_curves}.
The experimental relaxation data of the nuclear magnetization component $M_\mathrm{z}$ were modeled as 
\begin{equation}
\label{eq:T1}
M_\mathrm{z} \left( \tau_1 \right)=M_0 \left[ 1-2 \exp\left( -\left( \tau_1 /T_1 \right)^{\beta} \right) \right],
\end{equation}
where $\beta$ is a stretching exponent.
The temperature dependence of the stretching exponent $\beta$ at 7~T is shown in Fig.~\ref{fig:1overT1_and_beta}(b).
We find that the stretching exponent is essentially temperature independent in the investigated temperature range, with experimental values close to unity, indicating a uniform spin-lattice relaxation rate of the $^{31}$P nuclear ensemble.

\begin{figure}[!ht]
	\centering
	\includegraphics[width=1\linewidth]{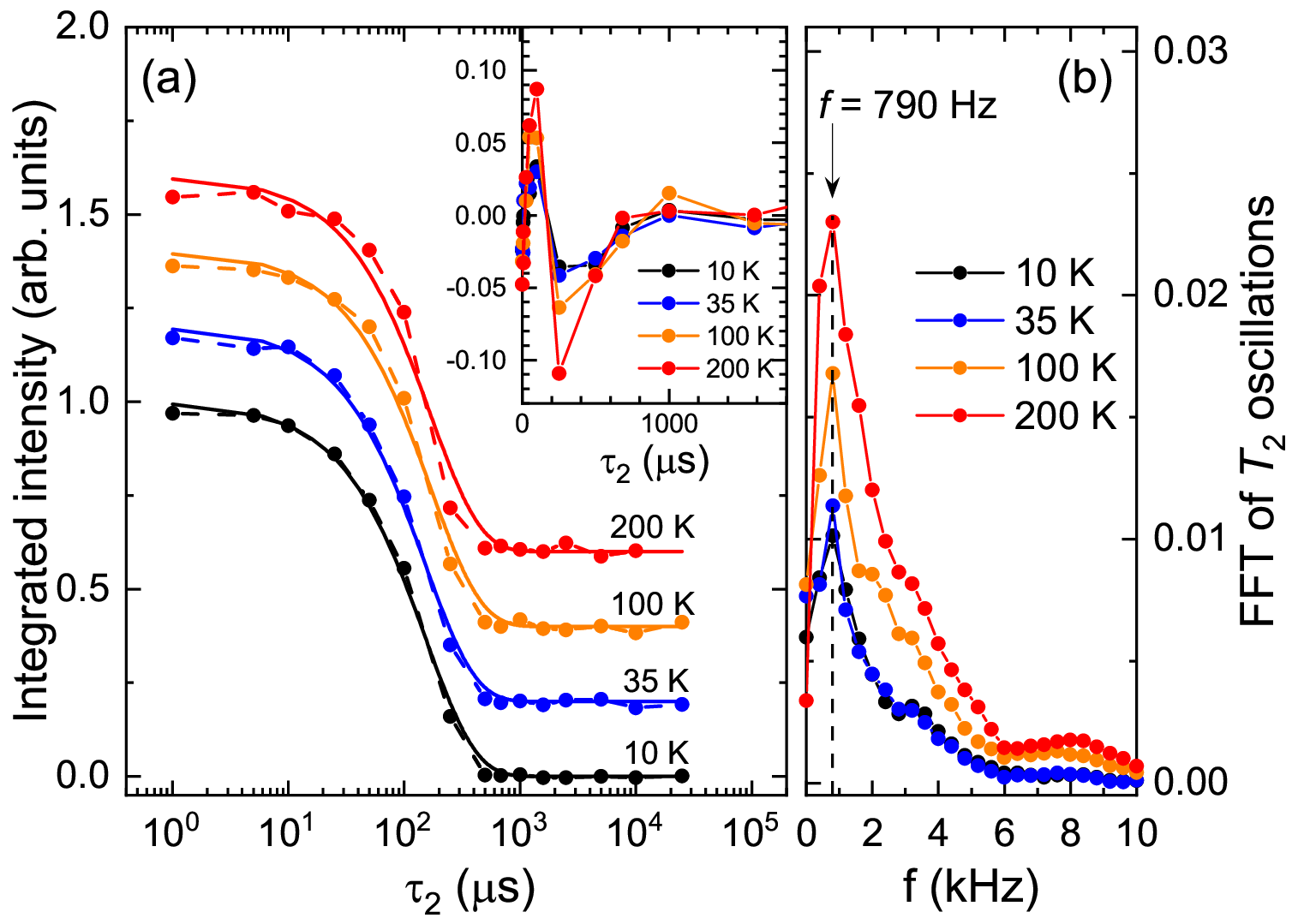}
	\caption{(a) Experimental relaxation data of the $^{31}$P nuclear magnetization component $M_\mathrm{xy}$ (circles connected by dashed lines) at 10, 35, 100, and 200 K, modeled by Eq.~(\ref{eq:T2}) (solid lines). The vertical scale corresponds to the data recorded at 10~K. The curves at higher temperatures are shifted vertically for clarity. The inset shows the deviation of the experimental data from a single-exponential decay. (b) Fast Fourier transform of the oscillatory component of the $T_2$ relaxation at 10, 35, 100, and 200~K. The vertical dashed line indicates a frequency of 790(70)~Hz, corresponding to the maximum amplitude of the FFT spectra.}
	\label{fig:T2_curves}
\end{figure}

\textit{The nuclear spin-spin relaxation time $T_2$}. The nuclear spin-spin relaxation time $T_2$ was recorded with a Hahn spin-echo pulse sequence with a typical $\pi/2$ pulse duration of 2.5~$\mu$s and an output power of 30~W.
Typical relaxation curves recorded at 10, 35, 100, and 200 K are shown in Fig.~\ref{fig:T2_curves}(a).

The exponential $T_2$ relaxation is modulated by a weak oscillatory component, stemming from the indirect spin-spin coupling between $^{31}$P and $^{19}$F nuclei.
The deviations of the experimental data from a single-exponential decay of the form
\begin{equation}
\label{eq:T2}
M_\mathrm{xy} \left( \tau_2 \right) = M_0 \exp(-2 \tau_2 / T_2)
\end{equation}
is presented in the inset of Fig.~\ref{fig:T2_curves}(a).
The oscillatory component yields a frequency of about 790(70)~Hz, see the fast Fourier transforms (FFT) of the oscillatory components at 10, 35, 100, and 200~K in Fig.~\ref{fig:T2_curves}(b).
The frequency of the oscillatory component is in good agreement with $J = 712$~Hz, determined by the MAS-NMR spectroscopy, see Fig.~2(b) in the main text.

\end{document}